\documentclass[twocolumn]{aastex631}
\usepackage{amsfonts,amsmath,graphicx,url,hyperref,nicefrac}
\usepackage{amsmath}
\usepackage{amssymb, verbatim, subfigure, paralist, soul, comment, multirow}
\usepackage{natbib}

\usepackage{color}
\shorttitle{GRB and Jet Engine Duration Relation} 
\shortauthors{Dastidar \& Duffell}

\begin{document}

\title{A Novel Relationship Between Gamma Ray Burst Duration And Photospheric Radius}

\author[0009-0000-6548-6177]{Ranadeep G. Dastidar}
\affiliation{Department of Physics and Astronomy, Purdue University, 525 Northwestern Avenue, West Lafayette, IN 47907, USA}

\author[0000-0001-7626-9629]{Paul C. Duffell}
\affiliation{Department of Physics and Astronomy, Purdue University, 525 Northwestern Avenue, West Lafayette, IN 47907, USA}

\begin{abstract}

Long Gamma Ray Bursts (lGRBs) are associated with jets in Type Ic broadline supernovae. The Collapsar model provides a theoretical framework for the jet formation from the core collapse of a massive star in such supernovae. The GRB can only be produced after a successful jet break out from the star. Under this formalism the GRB duration ($t_{\rm{90}}$) has been hypothesized to be the difference between the central engine activity duration ($t_{\rm{eng}}$) and the jet breakout time ($t_{\rm{bo}}$), that is $t_{\rm{90}} = t_{\rm{eng}} - t_{\rm{bo}}$. This disallows $t_{\rm{90}} > t_{\rm{eng}}$ and puts a lower bound on successful lGRB jet central engine duration ($t_{\rm{eng}} > t_{\rm{bo}}$), various numerical simulations have shown otherwise. This study considers a photospheric GRB emission from a relativistic jet punching out of a Wolf-Rayet-like star. We use the bolometric lightcurve generated to calculate the lGRB duration ($t_{\rm{90}}$) for varying engine duration. We find for longer engine duration the lGRB lightcurve reflects the jet profile and $t_{\rm{90}} \approx t_{\rm{eng}}$. While for shorter engine duration, the $t_{\rm{90}}$ has photospheric radius ($R_{\rm{ph}}$) dependence. This can be modeled by a relation, $t_{\rm{90}} = t^{\rm{90}}_{\rm{eng}} +  0.03\left(\frac{R_{\rm{ph}}}{c}\right)$, where c is the speed of light, with a lower bound on $t_{\rm{90}}$ for a successful lGRB. This relation should be most relevant for possible low-luminous lGRBs originating from a collapsar with central engine duration comparable to the jet breakout time. 

\end{abstract}


\section{Introduction} \label{sec:intro}

Gamma Ray Bursts (GRB) are typically classified by their temporal prompt emission phase and marked by excessive sub-MeV or higher energy photons. The burst duration is defined by the time taken between the 5\% to 95\% of the total fluence of the burst, called $t_{\rm{90}}$. A population distribution plot for the number of GRBs and their duration from BATSE observations produces a bimodal distribution. This bimodal Gaussian distribution hints at two classes of GRBs separated around $t_{\rm{90}} \sim 2$ s margin. These two separate classes are identified as short-Gamma Ray Burt (sGRB) and long-Gamma Ray Burt (lGRB) \citep{Kouveliotou+1993}. The two burst classes have since been associated with separate astrophysical processes. The sGRB has been associated with Neutron Star - Neutron Star merger \citep{Eichler+1989}. The lGRB are associated with collapsing massive stars and confirmed with direct observations \citep{Galama+1998}. Several authors have also hinted towards a third classification of GRBs based on $t_{\rm{90}}$ population distribution \citep{Horvath+1998}, as a hidden population of intermediate or soft GRBs \citep{Mukherjee+1998} or low luminous GRBs \citep{Bromberg+2011}. For this study, we investigate lGRBs from collapsing massive stars.

The Collapsar Model \citep{Woosley+1993, MacFadyen+1999} provides a theoretical framework for Gamma Ray Bursts from the core collapse of a supermassive star. According to the Collapsar Model, the stellar interior collapses on itself, leaving behind a central compact object. The compact object accretes the remaining collapsed material, and this accretion power is converted into a relativistic jet. The jet punches through the star, and under the right conditions breaks out of the stellar surface. The collapsar model for GRB prompt emission has been numerically studied intensively (\cite{James+2022, Gottlieb+2023, Hamidani+2025} and others). Not all core collapse supernovae lead to a GRB. It is argued that such jets are choked before punching out of the star with the energy deposited in the jet cocoons \citep{Lazzati+2011, Piran+2019}. Jets with engine duration ($t_{\rm{eng}}$) less than the jet breakout time ($t_{\rm{bo}}$) are expected to fall under this category (i.e. $t_{\rm{eng}} < t_{\rm{bo}}$). For $t_{\rm{eng}} > t_{\rm{bo}}$, due to relativistic effects, the observed duration of the GRB ($t_{\rm{90}}$) reflects the jet operation time or equivalently the engine duration \citep{Sari+1997}. The jet, central engine, and collapsar interaction effects are imprinted on the GRB prompt emission \citep{Lazar+2009}. This leads to the well known lGRB duration relation $t_{\rm{90}} = t_{\rm{eng}} - t_{\rm{bo}}$ \citep{Bromberg+2012}. The GRB-engine duration relation, under a given universal breakout time and power-law distribution in $t_{\rm{eng}}$ can be used to model the observed lGRB population distribution \citep{Bromberg+2012, Sobacchi+2017, Maria+2017}. Using jet models in which engine power and activity time-scales are independent of each other, \citep{Petropoulou+2020} found that the success of jet breakouts and detectability of lGRBs can be accounted for by the combined effects of luminosity and jet duration. The engine duration and luminosity are also independent of each other in our jet model. In this study, we explore the effects of varying jet duration while keeping luminosity fixed across all jet models. The effects of varying luminosity will be explored in future studies.

However, low-luminous lGRBs (llGRBs) do not fit the population predicted by the above relation \citep{Bromberg+2011}.  Shock breakout of the jet from the star has been proposed as a natural explanation for low luminous GRBs \citep{Nakar+2012}.  Moreover, jets with $t_{\rm{eng}} \sim t_{\rm{bo}}$ are expected to lead to llGRBs and therefore these bursts should be the most direct probe to the $t_{90}$ correlation.  The low duration lGRB population distribution shows a plateau-like feature, but this could potentially be attributed to a purely statistical and sample-incompleteness interpretation \citep{Osborne+2024}. Additionally, \cite{Barnes+2018} found $t_{\rm{90}}$ a factor of 5 greater than the $t_{\rm{eng}}$ for engine duration less than the jet break out time for a Type Ic broad-line supernova. These results motivate a closer look at the lGRB duration and engine duration correlation, especially for shorter engine duration.

In this work, we study the jet launched by a central engine due to the core collapse of a Wolf Rayet star. The jet travels through the star and breaks out into a circumstellar wind medium. The feasibility of a successful long Gamma Ray Burst is checked for a varying central engine activity duration ($t_{\rm{eng}}$). We then investigate imprints of the central engine activity duration on the lGRB prompt emission duration ($t_{\rm{90}}$). This is done assuming a photospheric GRB emission \citep{Meszaros+2000, Lazzati+2013, Ito+2015, Gottlieb+2019} model using bolometric luminosity for the prompt emission lightcurve. We further study the effects of different photospheric radii ($R_{\rm{ph}}$) on the lightcurve and hence the burst duration ($t_{\rm{90}})$.

This paper is arranged as follows. Sec. \ref{sec:NumSetUp} discusses the motivation and numerical setup for our jet-star interaction model and bolometric luminosity calculations for a photospheric Gamma Ray Burst. This is followed by a detailed discussion of the hydrodynamical evolution of the jet-star system in Sec. \ref{sec:Results}. In this section, we also report the results of our model. Sec. \ref{sec:Discussion} analyses and discusses these results. Here we propose a novel relationship between lGRB burst duration and the photospheric radius. Finally, we present the conclusions of our findings and their implications in Sec. \ref{sec:Conclusion}.

\section{Numerical Setup} \label{sec:NumSetUp}

Our stellar progenitor model is motivated by \cite{Barnes+2018}. The innermost region of a WR star is assumed to have collapsed into a compact object. A stellar model for a WR star is inducted for the outer layers. This assumption works since the outer layers of the star are unaffected by the core collapse and insensitive to the density profile of the collapsed region. These assumptions may affect the estimations of the elements in the supernova phase, but lGRB dynamics and lightcurve are not likely to be affected. Owing to the stripped-envelope nature of WR stars, a very low-density stellar wind density profile is used outside the star.

The \textit{Jet Code} is a 2D axisymmetric moving mesh relativistic hydrodynamics solver. As a result, the inner and outer boundaries move relative to the boundary velocities. Hence, the initial grid resolution is overall conserved over the entire temporal domain. The initial grid resolution is set to $\delta \theta/\theta \approx 1.9\times10^{-3}$. The hydrodynamic jet-star evolution starts at 0.01 sec, when the jet is seeded within the cavity, and evolves until much later into the jet deceleration phase.

\begin{figure}[h!]
\centering
\includegraphics[width=0.25\textwidth]{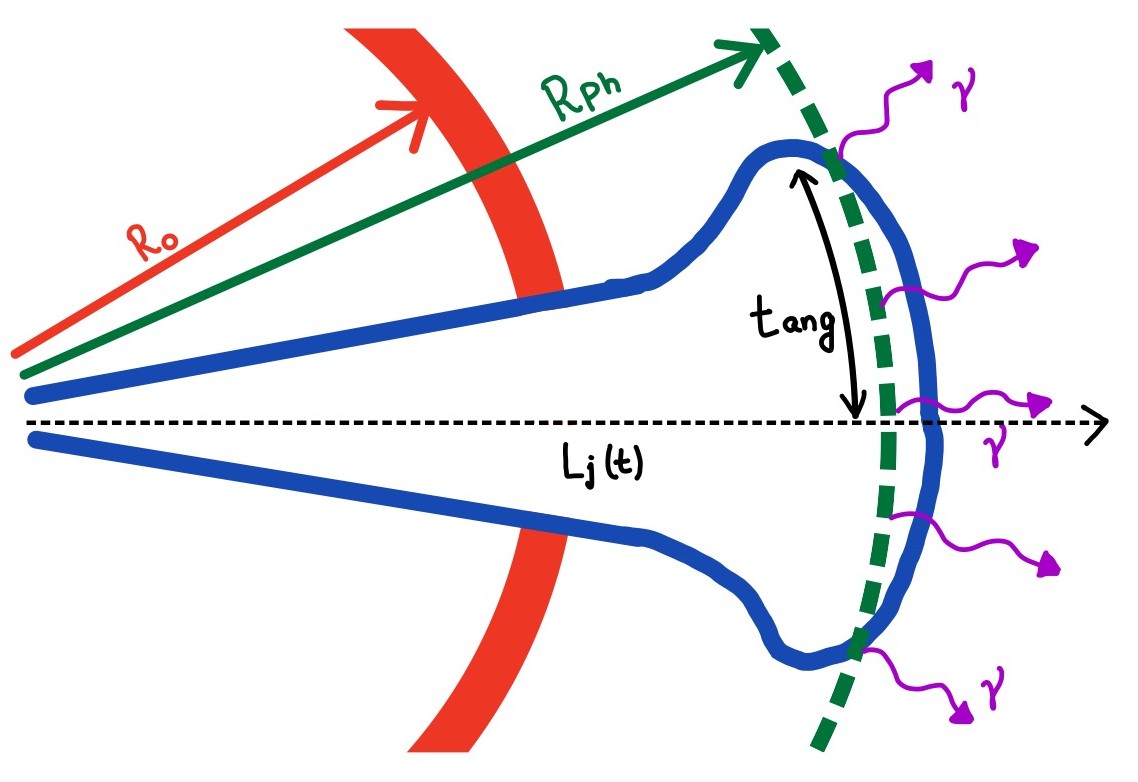}
\caption{\footnotesize{Pictorial depiction of the variables involved in lGRB prompt emission. The jet (blue border), with an input luminosity profile ($L_j(t)$), punches out of the stellar surface (red border) at $R_{\rm{wr}}$}. Upon crossing the photosphere (Green dashed, $R_{\rm{ph}}$), we estimate the bolometric luminosity for lGRB prompt emissions ($\gamma$-rays) from the relativistic material in the jet. The black dashed arrow represents the jet axis (observer's line of sight). We also correct for the photon arrival delay from off-axis $\gamma$-rays by adding the angular delay time, $t_{\rm{ang}}$ Eq. (\ref{Eq:t_ang}).}
\label{fig:tang_cartoon}
\end{figure}

Inside the star, the collapsed region (or cavity) is modeled by removing material from the computational domain inner to $R_{\rm{cav}} = 1.5\times 10^{-3} R_{\odot} \approx 1000 \text{km}$. The density in this region is set to $10^{-3}$ times the density at $R_{\rm{cav}}$. The density beyond the cavity ($r > R_{\rm{cav}}$) is set to the pre-collapse density profile of a WR star. The density profile ($\rho_{\rm{init}}$) is given by :

\begin{equation}
    \label{Eq:Density_WR}
    \rho_{\rm{init}} = \frac{0.0615 M_{\rm{wr}}}{R_{\rm{wr}}^3}\left(\frac{R_{\rm{wr}}}{r}\right)^{2.65}\left(1-\frac{r}{R_{\rm{wr}}}\right)^{3.5} 
\end{equation}

Where the stellar radius $R_{\rm{wr}} = 1.6R_{\odot}$ is the extent of the stellar atmosphere, and $M_{\rm{wr}} = 2.5M_{\odot}$ is the stellar mass beyond the cavity ($M_{\rm{cav}} \approx 1.4M_{\odot}$). These values correspond to a total stellar mass $M \lesssim 4 M_{\odot}$. A detailed description of the model is presented \cite{Barnes+2018}. Following the core collapse, the remaining stellar material is continuously in a free-fall state. Our hydrodynamic simulation does not include gravity and therefore does not account for free-falling material dynamically. However, we do not expect the innermost stellar material to have a significant impact on the final result.

The central compact object after the core collapse plays the role of a jet launching engine in our model. In essence, we do not need an exact structure of the compact object. Rather, it is modeled as a jet nozzle with an opening angle, $\theta_0$, injecting energy at a constant luminosity, $L_0$. A Gaussian angular structure is used for the jet. The jet and its counter jet are launched from the central engine along the polar axis. The jet is launched deep within the stellar cavity, at a radial distance of 150 kms from the center. The engine has a baryon loading ratio of energy-to-mass set to 1000.0. Though the jet typically entrains enough mass on its way out to make the final value lower. The temporal cut-off for the modeled jet depends on the jet launching mechanism. Typically, a sharp jet cut-off model is implemented (Eq. \ref{Eq:sharp_cutoff}) (\cite{Gottlieb+2024}). However, the exact time dependence is unknown and may not be the same for every GRB as light curves are diverse (\cite{Young+2003}). Hence, along with a sharp engine cut-off model, we also consider two other jet cut-off models. One where the engine injects energy which exponentially decays over a characteristic timescale, ``$t_{\rm{eng}}$" (Eq. \ref{Eq:exp_cutoff}) (\cite{Barnes+2018}). The results are primarily discussed based on this model. Other, where the engine decays as $1/t^2$ following equation Eq. \ref{Eq:square_cutoff} (\cite{Duffell+2018, Zuo+2025}). A lGRB is typically expected to have the same duration as this time scale for the sharp cut-off model. While for the later models, the engine persists longer than “$t_{\rm{eng}}$”, the corresponding “$t_{90}^{\rm{eng}}$” can be calculated by integrating the luminosity for 90\% of the total energy. This gives the intrinsic $t_{90}$ for the models in Eq. \ref{Eq:cutoffs} as; $0.9t_{\rm{eng}}$, $ln(10)t_{\rm{eng}}$ and $9t_{\rm{eng}}$ correspondingly. In this study, we explore this correlation between $t_{\rm{eng}}$ and $t_{\rm{90}}$ for varying engine activity time.

\begin{subequations}
\label{Eq:cutoffs}
\begin{align}
L_{\rm eng}(t) &=
\begin{cases}
    L_0, & t \leq t_{\rm eng}, \\
    0,   & t > t_{\rm eng},
\end{cases}
\label{Eq:sharp_cutoff} \\[6pt]
L_{\rm eng}(t) &= L_0 \exp\left(-\frac{t}{t_{\rm eng}}\right)
\label{Eq:exp_cutoff} \\[6pt]
L_{\rm eng}(t) &= \frac{L_0}{\bigl(1 + t/t_{\rm eng}\bigr)^2}
\label{Eq:square_cutoff}
\end{align}
\end{subequations}

We focus on a single maximum luminosity for all of our simulations. That is, we fix $L_0 = 1.63\times 10^{52}$ erg/s and vary $t_{\rm{eng}}$ over a range (see Sec. \ref{sec:Results}). This results in jets with different total energy associated with each engine duration. For a given $t_{\rm{eng}}$, the total engine energy $E_{\rm{eng}} = L_0t_{\rm{eng}}$, with half the energy in each of the jets and the counter jet. The engine opening angle is fixed at $\theta_0 = 0.2 \mbox{ rad }( = 11.5^\circ)$. For a detailed description of the jet nozzle refer to \cite{Duffell+2015}.

\begin{figure*}
\centering
\includegraphics[width=0.8\linewidth]{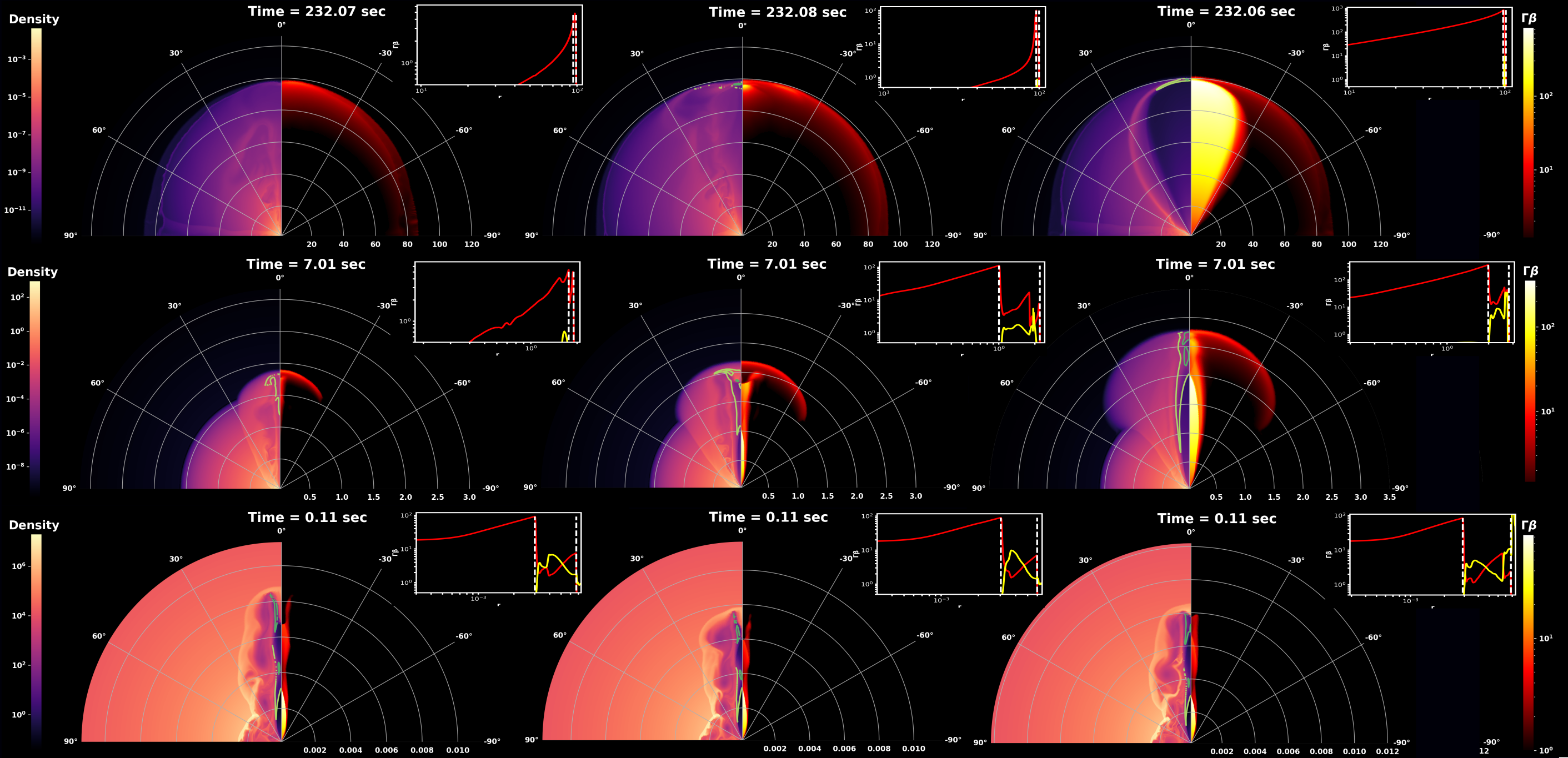}
\caption{\footnotesize{Snapshots of the hydrodynamical evolution of our model. This figure is divided into 9 panels. Each column of 3 vertical panels represents a single engine duration plotted at different evolution times. With each panel, the line plot shows the four velocity (red) and luminosity (yellow) along the jet core. The white dashed lines enclose the jet head ($dR_{uu}$). The left column plots for $t_{\rm{eng}} \approx 0.7 \mbox{ sec} (\ll t_{\rm{bo}})$, $t_{\rm{eng}} \approx 3.2 \mbox{ sec} (\sim t_{\rm{bo}})$ in the center column, and $t_{\rm{eng}} \approx 232 \mbox{ sec} (\gg t_{\rm{bo}})$ for the right column. Each row of 3 horizontal panels plots the jet evolution after the same time, for different $t_{\rm{eng}}$. The bottom panel shows the jet while it is still within the stellar envelope (the extent of the envelope is $R_{\rm{wr}} = 1.6$ radial code units). The middle panel shows the jet cocoon soon after the breakout, and the top panel is plotted when the jet reaches $R_{\rm{ph}} = 100 R_{\rm{wr}}$. In each panel, the right quadrant plots the density distribution, while the left quadrant plots the four velocity distribution (with cut-off, $\Gamma\beta > 1$). The most luminous region that emits the lGRB has been shown in green contours in each panel. For $t_{\rm{eng}} \ll t_{\rm{bo}}$, the jet slows down rapidly, and we see no lGRB contours in the top panel. For $t_{\rm{eng}} \sim t_{\rm{bo}}$ at all times (all $R_{\rm{ph}}$), and $t_{\rm{eng}} \gg t_{\rm{bo}}$ at earlier times (smaller $R_{\rm{ph}}$), the lGRB is emitted from the jet head and hence their $t_{90}$ show photospheric dependence. Finally, for $t_{\rm{eng}} \ll t_{\rm{bo}}$, the lGRB is dominated by the shocked jet, and for most photospheric radii we get $t_{90} \sim t^{90}_{\rm{eng}}$.}}
\label{fig:hydro_evol}
\end{figure*}

\subsection{Prompt Emission}

The exact mechanism behind the prompt emission in a lGRB is still an area of active debate. However, it is universally accepted that a fraction of the internal kinetic energy of the jet or the jet star interacted material is converted to the prompt emission via some internal dissipation mechanism \citep{Zhang+2006}. This energy dissipation can be due to external shock interactions \citep{Rees+1992} or internal dissipation shocks \citep{Narayan+1992}. However, external shocks are strongly disfavored for explaining prompt GRB emission \citep{Sari+1997}. The existence of thermal emission in lGRB has been shown in observations \citep{Peer+2015} and spectral properties \citep{Axelsson+2015}. This motivates a photospheric dissipation model (see \cite{Bing+2018} for a detailed list). By “photosphere" here we refer to the radius at which photons last scatter before leaving the jet to give rise to the prompt lGRB. Hence, the photosphere is technically the photon diffusion surface. In this work, we will simply refer to this as the “photospheric radius" ($R_{\rm{ph}}$). A caveat to the photospheric emission model is that it cannot account for the X-ray steep decay in the end of the prompt emission phase, which is better explained by the internal shocks \citep{Hascoet+2012}. Purely thermalized jet with photospheric emission fails to produce the observed broad range of peak photon energy $E_{\rm{p}}$, instead a synchrotron photon source is needed at a smaller radius to explain higher $E_{\rm{p}}$ \citep{Vurm+2013}. While the radiation mechanism for GRBs is not yet settled, observations hint at a combined thermal and non-thermal mechanism \citep{Oganesyan+2017, Ravasio+2019}. \cite{Paz+2017} have shown that prompt emission from Comptonization occurs at smaller radii, while synchrotron emission dominates at larger radii. The thermal component can be interpreted as the jet's photospheric emission \citep{Guiriec+2017} and is the focus of this study. In this study, we explore the temporal structure of prompt emission against light crossing time within the jet's internal radial structure while remaining agnostic to the radiation mechanism. Within this framework, we do not explore the spectral properties of the GRB and hence assume a bolometric and purely thermal emission. Follow up of this work would include synchrotron emission and a study into the value of $E_{\rm{p}}$ for the bursts.

In this work, we look at the prompt emission duration ($t_{\rm{90}}$) for a lGRB due to photospheric emission. This is estimated from the bolometric lGRB lightcurve. Which in turn reflects the energy distribution in the jet-cocoon. We can therefore approximate the exact underlying emission mechanism for the lGRB as a photospheric emission from the dissipation of the internal jet energy. The exact value of the observed luminosity will scale with the energy conversion efficiency at the photospheric radius. We do not however expect this to modify the observed time scales for the prompt emission. The time scales will rather be dominated by the hydrodynamical timescales of the problem. Under these assumptions, we calculate the bolometric luminosity of the prompt emission by converting all of the internal energy of the jet material, after its interaction with the stellar envelope, as it crosses the photosphere. The observed luminosity is given by : 

\begin{equation}
    \label{Eq:Enthalpy}
    L(t) = 4\pi R_{\rm{ph}}^2\rho(t) h(t) \Gamma(t)^2\beta(t) c
\end{equation}

Where, $\rho(t)$ is the local density of the relativistic material at the photosphere ($R_{\rm{ph}}$) moving with Lorentz factor $\Gamma(t)$ and velocity $\beta c$, and $h$ is the specific enthalpy of the material. Since the production of $\gamma$-rays necessitates a relativistic boost, we consider only the material having $\Gamma\beta > 1$ for luminosity calculations. We use the ultra-relativistic equation of state $e = P/3$ ($e$ and $P$ being the internal energy density and pressure respectively) and $h = c^2 + 4P/\rho$. The luminosity calculated here assumes a complete conversion of the enthalpy to radiation. We use a non-dissipative photospheric emission mechanism. The photons are thermalized deep inside the jet and escape as and when they reach the photospheric radius. This overestimates the luminosity values in the lightcurve, but is intended to give a rough estimate of the time dependence of the bolometric luminosity.

To account for the emission contributions from the material directed away from the line of sight axis of the observer, an angular spreading time ($t_{\rm{ang}}$) is added to the observer time ($t_{\rm{obs}}$) as, $t_{\rm{obs}} = t + t_{\rm{ang}}$ (Fig. \ref{fig:tang_cartoon}). Where $t$ is the local time in the blast frame. Since we consider a photosphere at rest, and not co-moving with the jet, the photon arrival time is intrinsically included in $t_{\rm{obs}}$ calculation.

\begin{equation}
    t_{\rm{ang}} = \frac{R_{\rm{ph}}}{c}\left[1-\rm{cos}\left(\frac{1}{\Gamma}\right)\right]
    \label{Eq:t_ang}
\end{equation}

The hydrodynamical quantities are obtained from the numerical simulations from \textit{the Jet Code} and the bolometric luminosity is calculated with \textit{the Firefly Code} \citep{Dastidar+2024}.

\section{Results} \label{sec:Results}

The jet propagation through the WR star in our simulations is shown in Fig. \ref{fig:hydro_evol}. Energy is dumped at the center of the collapsed star through the nozzle as a function of time (Eq. \ref{Eq:cutoffs}). We computed the evolution for a range of engine durations from $0.2 \mbox{ s} < t_{\rm{eng}} < 465 \mbox{ s}$. For a given simulation (single $t_{\rm{eng}}$), the jet hence formed travels through the cavity and enters the stellar envelope. Photospheric emissions from the relativistic materials then emit a prompt GRB. In this section, we first discuss the jet propagation through the star for a typical engine duration, $t_{\rm{eng}} > t_{\rm{bo}}$. Then we present our results on the effects of the different engine durations on these dynamics, followed by the signatures it leaves behind in the GRB lightcurve and duration.

\subsection{Jet Propagation}

\begin{figure*}
\centering
\includegraphics[width=0.9\linewidth]{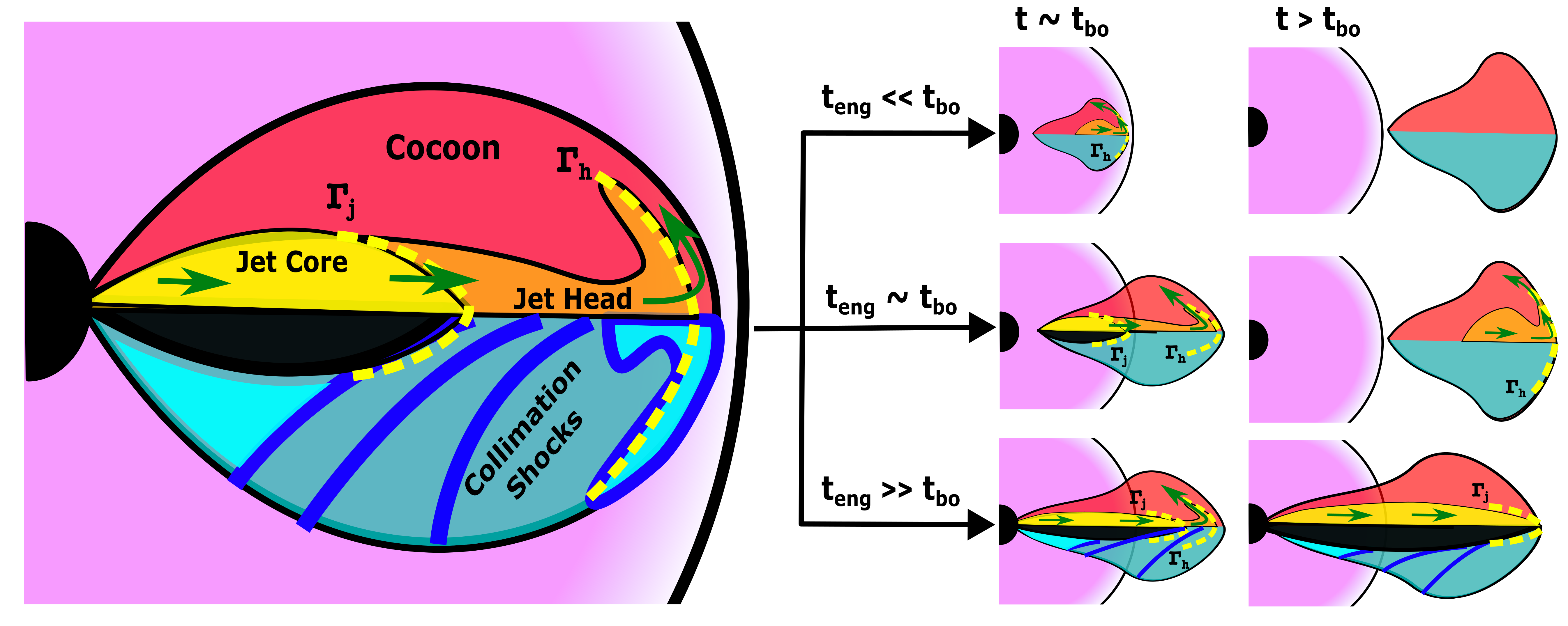}
\caption{\footnotesize{A schematic diagram for the jet propagation. The left image shows the formation of the slower, massive jet head (orange) and the faster, low-density jet core (yellow). The jet head consists of shocked jet and stellar, while the jet core is fed directly by the central compact object. Both of them drive a shock in the front, marked by $\Gamma_h$ and $\Gamma_j$ respectively. Depending on the central engine activity duration, the jet either shuts off before it breaks out of the star ($t_{\rm{eng}} \ll t_{\rm{bo}}$, top), shuts off near around breakout ($t_{\rm{eng}} \sim t_{\rm{bo}}$, middle) or continues long after the break out ($t_{\rm{eng}} \gg t_{\rm{bo}}$, bottom). Following the breakout, the jet either consists of just the jet cocoon (with all the jet energy transferred to the cocoon), or an elongated jet head (as the jet core transfers all its energy to the jet head), or the jet core (still being fed by the central compact object) surrounded by the cocoon, correspondingly. This is also evident in the simulated outputs shown in Fig. \ref{fig:hydro_evol}.}}
\label{fig:jet_evol_cartoon}
\end{figure*}

The jet forms the typical jet-cocoon structure \citep{Bromberg+2011th} quickly after it is launched from the cavity into the stellar envelope. After the jet is launched from the collapsed stellar core, the high velocities at the front of the jet drive a relativistic shock as it interacts with the stellar envelope. We refer to this as the leading shock ($\Gamma_h$). This slows the jet's front while accumulating stellar material. While the jet is continuously fed by the central engine, it drives a second faster shock into this accumulated material within the jet. We refer to this shock as the trailing shock ($\Gamma_j)$. The part of the jet in between these shocks, composed of both the slower-moving jet and accumulated stellar material, is called the jet head. Any material entering this region is already shocked, hence it is also referred to as the shocked jet head. The part of the jet behind the trailing shock is coming directly from the central engine and is therefore faster and unshocked. This region of the jet is called the unshocked jet. Material also continues to leak out of the jet head via the leading shock. These materials are slower than the jet and sub-relativistic, forming a balloon-like structure on both sides of the jets. This is called the cocoon, and together with the jet, it is referred to as the jet-cocoon structure. Fig. \ref{fig:jet_evol_cartoon} shows a pictorial representation of these regions. The jet head is marked by high pressure and sub-relativistic velocity. Most of the energy is yet to be injected into the jet. The cocoon is sub-relativistic and has isotropic pressure which collimates the jet. The shocked collimated jet is ultra-relativistic and at pressure equilibrium with the cocoon. The unshocked jet is a very small conical structure at the center of the domain. This results in two relativistic shock fronts moving forward with different Lorentz factors. The leading shock (the shocked jet front) has a lower Lorentz factor than the trailing shock (the unshocked jet front). Similar results have been seen by \cite{Barnes+2018}.

As the jet propagates, energy is continuously injected into it, which along with the increased ram pressure from collimation pushes the jet head to higher velocities. This increases the cocoon height. The injected jet profile is still reflected in the unshocked jet. The cocoon elongation increases the jet head height. Almost all the energy injected in the jet is confined within this jet head, as it is the most dense region. In our simulations, all the jets (with different $t_{\rm{eng}}$) reach this phase of the evolution. The jet at this stage is confined well within the star in a cocoon at a higher pressure than the ambient.

For engines that shut off at this stage, which corresponds to $t_{\rm{eng}} \ll t_{\rm{bo}}$, the unshocked jet transfers all its remaining energy to the jet head while the jet remains collimated. The double shock structure within the jet vanishes at this point. The collimated jet-cocoon structure then travels through the stellar envelope. As it breaks out of the star, the cocoon experiences a sudden decline in ambient density and undergoes free expansion. This is marked by a small interval of shock acceleration for the jet head ($\Gamma\beta \sim 4-40$). Soon the jet transfers its energy to the cocoon as it expands and becomes spherical. The jet head starts to decelerate during this period.

If the engine continues it increases the jet head height, which expands the cocoon and decreases its pressure. The cocoon pressure eventually becomes insufficient to collimate the jet. As a result, the jet head speeds up to relativistic velocities. This causally disconnects parts of the jet.

If the engine deposits still more energy into the jet, the jet becomes uncollimated. The unshocked jet accelerates through the shocked jet, approaching the jet head. Since the unshocked jet front is much faster than the shocked jet, most of the energy is stored in the now relativistic jet head, which gradually seeps into the cocoon via a thin layer from the jet head. Two distinctive relativistic shock fronts mark the shocked and unshocked regions of the jet.
At this stage, the jet-cocoon structure is uncollimated with its head close to the stellar surface.

If the engine is turned off at this stage ($t_{\rm{eng}} \sim t_{\rm{bo}}$), most of the energy is stored in the jet head, while the fastest region is still the unshocked jet. As there is no further injection of energy into the system, the uncollimated jet gradually spreads out, increasing the energy transfer rate into the cocoon. The unshocked jet slows down as it transfers all its energy to the jet head and eventually dies out. The shocked jet initially gains this velocity before slowing down to sub-relativistic velocity after it loses all its energy to the cocoon. The cocoon balances out the pressure and becomes spherical. In this scenario, there are no internal collisions, either between the two shock fronts within the jet (unshocked and shocked jet) or the shocked jet front and the jet head.

\begin{figure}[h!]
\flushleft
%\hspace{-2cm}
\gridline{\fig{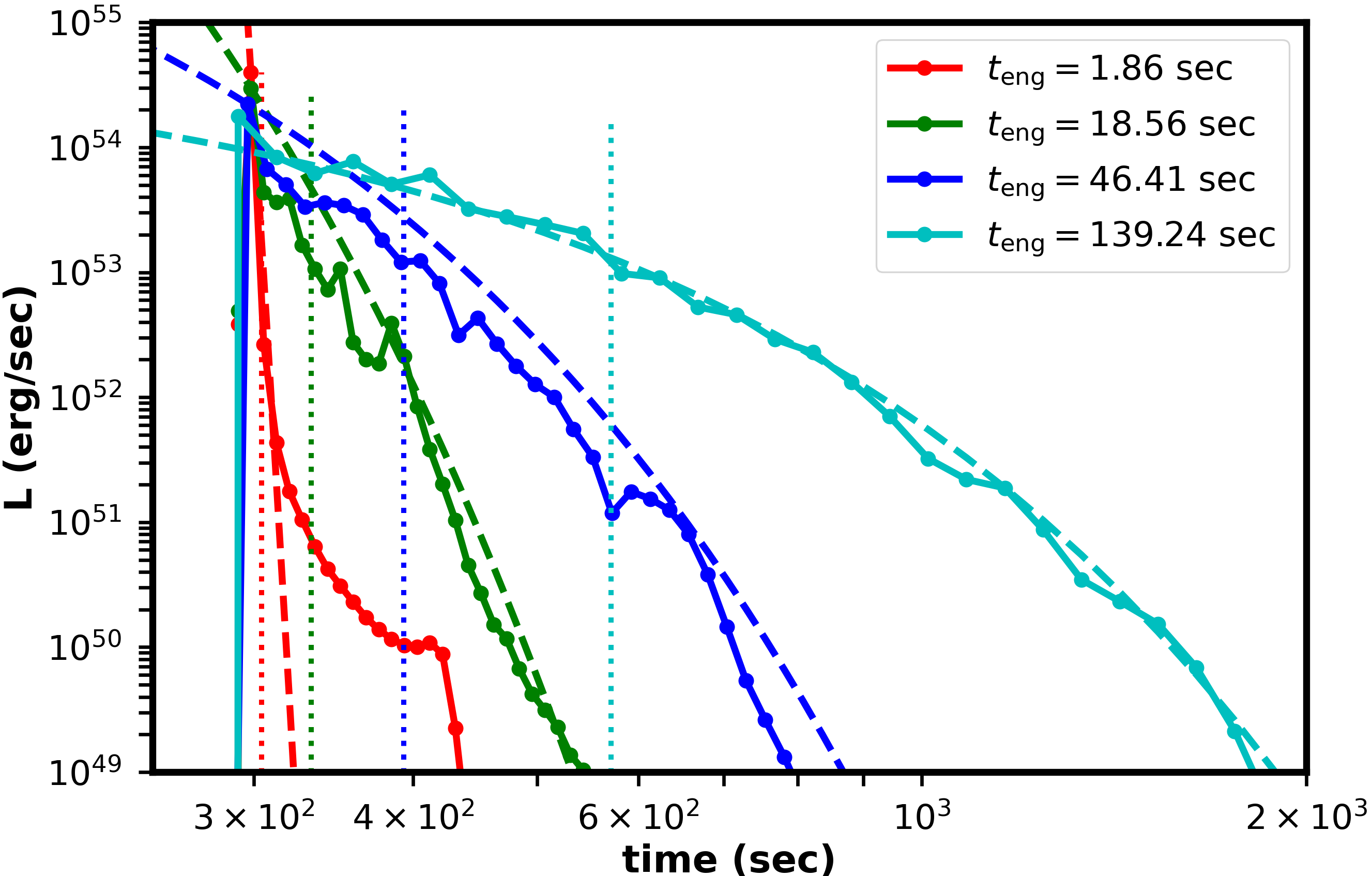}{0.75\linewidth}{\footnotesize{(a) lGRB lightcurve for various engine duration emitted at $R_{\rm{ph}} = 80 R_{\rm{wr}} \approx 8.9\times 10^{12}$ cm.}}}
    \gridline{\fig{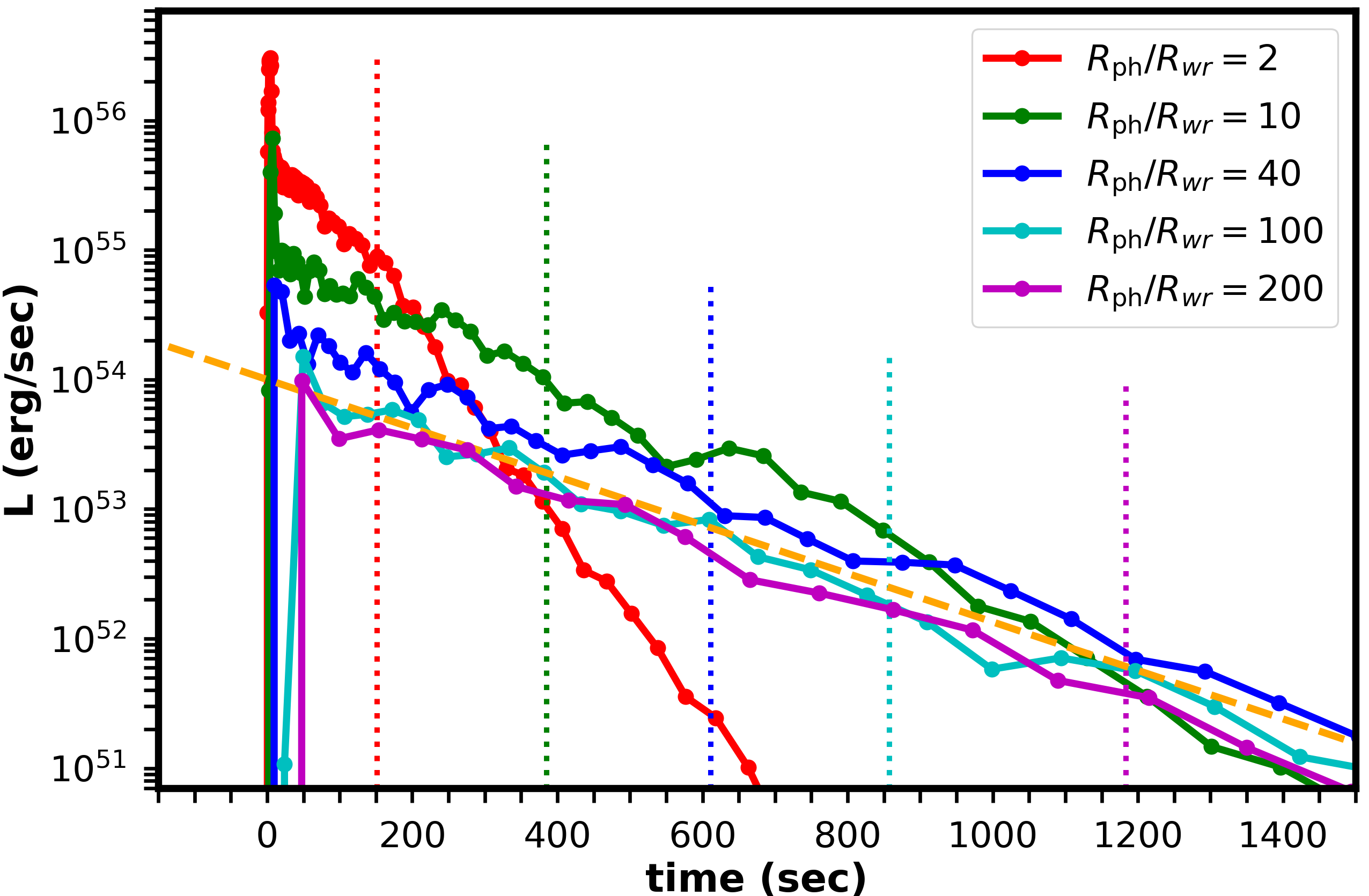}{0.75\linewidth}{\footnotesize{(b) lGRB lightcurve for $t_{\rm{eng}} \approx 232$ sec if emitted at different radii.}}}
    \caption{\footnotesize{lGRB lightcurve for our model. In (a), the coloured dashed lines represent the input jet profile for the corresponding $t_{\rm{eng}}$. While the vertical coloured dotted lines are the corresponding $t_{\rm{90}}$ (for both plots, (a) and (b)). For shorter engine duration (red, $t_{\rm{eng}} \sim 2 \mbox{ sec} < t_{\rm{bo}}$) the luminosity profile deviates from the input jet profile. While for longer engines (cyan, $t_{\rm{eng}} \sim 140 \mbox{ sec} > t_{\rm{bo}}$) the lightcurve reflects the input luminosity profile for the jet very well, hence we get $t_{\rm{90}} \sim t^{90}_{\rm{eng}}$. In (b) we adjust the time axis to coincide the multiple bursts at different $R_{\rm{ph}}$ for a direct comparison. The orange dashed line represents the input jet luminosity profile, $L_j(t) = L_0e^{-t/t_{\rm{eng}}}$. For the photosphere near the stellar surface (say, red curve) the lightcurve decays sharper than the input profile. While for $R_{\rm{ph}}$ further away they decay as the input profile, and their $t_{\rm{90}}$ saturates to $t^{90}_{\rm{eng}}$.}}
    \label{fig:lightcurves}
\end{figure}

\begin{figure*}
\centering
\includegraphics[width=0.65\linewidth]{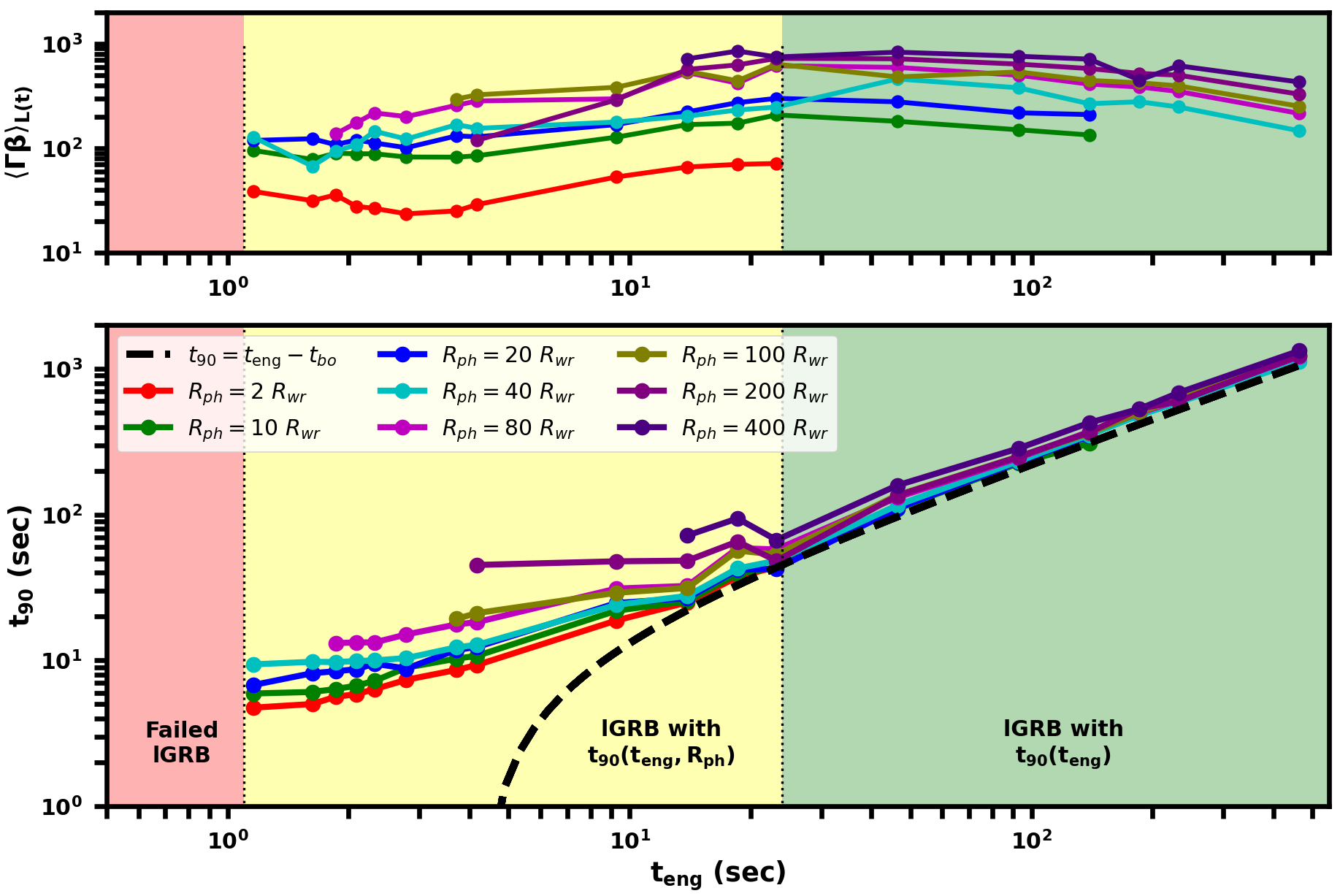}
\caption{\footnotesize{Plot for numerically estimated $t_{\rm{90}}$ for $t_{\rm{eng}}$. The top panel shows the luminosity averaged four velocity, $\left<\Gamma\beta\right>_{L(t)}$ for the all the $t_{\rm{eng}}$ modeled at different $R_{\rm{ph}}$. The lower panel plots the $t_{90} = t_{95} - t_5$, calculated from numerically generated lGRB lightcurves for the same. The jets having $\left<\Gamma\beta\right>_{L(t)} > 1$ emit lGRB. For $t_{\rm{eng}} < 1.2$ sec, we get $\left<\Gamma\beta\right>_{L(t)} < 1$. Hence the region marked red indicates lGRB less jets. The yellow region, spanning $1.2 \mbox{ sec} < t_{\rm{eng}} < 24 \mbox{ sec}$, indicates a rising $\left<\Gamma\beta\right>_{L(t)}$ (upper panel) and a strong photospheric dependence (lower panel). In the green region beyond $t_{\rm{eng}} > 24$ sec, $\left<\Gamma\beta\right>_{L(t)}$ saturates to $\sim 10^3$. The jets can not accelerate further due to baryon loading. Correspondingly in the lower panel, the lGRB duration is dictated by the central engine activity time. This gives, $t_{90} \sim t^{90}_{\rm{eng}}$. The black dashed line in the lower panel plots the Bromberg model for $t_{90}$ \citep{Bromberg+2012}. The solid black line marked, $t_{\rm{bo}}$ shows the breakout time for different $t_{\rm{eng}}$ obtained from numerical results.}}
\label{fig:t90_raw}
\end{figure*}

However, if the engine remains active and injects still more energy ($t_{\rm{eng}} \gg t_{\rm{bo}}$), the unshocked jet becomes faster. The cocoon loses more pressure and the jet becomes completely uncollimated. The unshocked jet front eventually collides with the shocked jet front. This eliminates shocked jet material along the jet axis. As a result, the hydrodynamic parameters behind the jet head exactly reflect the jet properties. Eventually, after the jet breaks out the jet head slows down, while the unshocked jet accelerates. The reverse shock in the jet head becomes weak and the unshocked jet front collides with the forward shock of the jet head. The jet then undergoes free expansion as a single shock front, unaffected by the ambient medium. Most energy is confined within a thin region behind the relativistic shock and in a thin layer around the jet.

\subsection{Prompt Emission}

As discussed earlier, we present our results for a photospheric prompt emission from the above jet dynamics for a WR star. The photosphere arises naturally when the total optical depth of the jet becomes unity. However, our codes do not intrinsically measure the opacity. Hence we consider artificially enforced photospheric radii in our models. To consider the effects of the photospheric radii we compute GRB bolometric lightcurves for each $t_{\rm{eng}}$ at 8 different radii, $R_{\rm{ph}}/R_{\rm{wr}} = [2,10,20,40,80,100,200,400]$. Although $R_{\rm{ph}}/R_{\rm{wr}}$ = 2 is very small for a realistic GRB emission site, it is included to obtain the appropriate limits to the correlation. We acknowledge that the photospheric radius is a function of the Lorentz factor and the luminosity, and hence arises naturally for a given set of jet injection parameters. But for the purposes of this study, which investigate the $t_{90}$ variability with the GRB emission site, we treat it as a free parameter.

For different regimes of $t_{\rm{eng}}$, the jet undergoes different stages of its evolution as it breaks out of the star and interacts with the ambient medium. lGRB bolometric lightcurves for different engine duration ($t_{\rm{eng}}$) at a given photospheric radius ($R_{\rm{ph}} = 100 R_{\rm{wr}} \approx 1.1\times 10^{13} \mbox{ cm}$) are plotted in Fig. \ref{fig:lightcurves}. Input analytical functions for the jet are plotted as dashed lines over the lightcurve. For shorter engine duration ($t_{\rm{eng}} \gtrsim t_{\rm{bo}}$), we see the lightcurve at $R_{\rm{ph}} = 100 R_{\rm{wr}}$ deviate significantly from the injected analytical jet temporal luminosity ($L(t)$). While for longer engine duration ($t_{\rm{eng}} \gg t_{\rm{bo}}$) the lightcurve reflects the jet structure exactly. We estimate the lGRB prompt emission duration ($t_{\rm{90}}$) from the bolometric lightcurve as the time it takes for the area under the curve to fall by 90\%. These are marked by the vertical dashed lines in Fig. \ref{fig:lightcurves}.

There is an important caveat to consider. Since our code does not assume a radiation mechanism, we set a cut-off for $\Gamma\beta > 1$ to be considered a GRB source. While the compactness argument hints towards a much faster Lorentz factor ($\Gamma > 100$) (\cite{Piran+2005}, and references therein), simulating such jets would require resolution much higher than we can generate. To accommodate this, we place the $\Gamma\beta > 1$ restriction, assuming the tail accelerates to a much higher $\Gamma$ not resolved by our simulations. Further, a jet breaking out of the star at relativistic velocities ($\Gamma\beta \geq 1$) may not have enough relativistic material to emit a GRB. To estimate this, we calculate the luminosity averaged four velocity ($\left<\Gamma\beta\right>_{L(t)}$) for each jet as it travels through the photosphere. The luminosity averaged four velocity is calculated as:

\begin{equation}
    \label{eq:lum_avg_four_velo}
    \left<\Gamma\beta\right>_{L(t)} = \frac{\int_{0}^{\infty} \Gamma\beta(t)\times L(t) dt}{\int_{0}^{\infty} L(t) dt}
\end{equation}

The results are plotted in Fig. \ref{fig:t90_raw}. The top panel plots the luminosity average four velocity for each jet having $\left<\Gamma\beta\right>_{L(t)} \geq 1$ at different photospheric radii. Only these engine durations emit a successful long gamma ray burst. We find that jets with engine duration less than $\sim 2$ seconds do not emit a GRB for our choice of jet and stellar parameters. While for longer engine duration ($t_{\rm{eng}} \geq 20$ seconds), $\left<\Gamma\beta\right>_{L(t)}$ saturates to a value of few thousands. The lower panel in Fig. \ref{fig:t90_raw} plots the $t_{\rm{90}}$ for the corresponding jets (the solid lines). We estimate $t_{\rm{90}}$ for only those jets which produce a lGRB.

We also argued in the previous section (\ref{sec:NumSetUp}) that $R_{\rm{wr}} \leq R_{\rm{GRB}} \leq R_{\rm{dec}}$. Hence, the shorter engine duration in Fig. \ref{fig:t90_raw} does not produce a lGRB at a larger photospheric radius. Therefore, we exclude the $t_{\rm{90}}-R_{\rm{ph}}$ combination; for jets that are decelerating at the given photospheric radius (that is, when $R_{\rm{dec}} < R_{\rm{ph}}$).

\begin{figure}
\centering
\includegraphics[width=\linewidth]{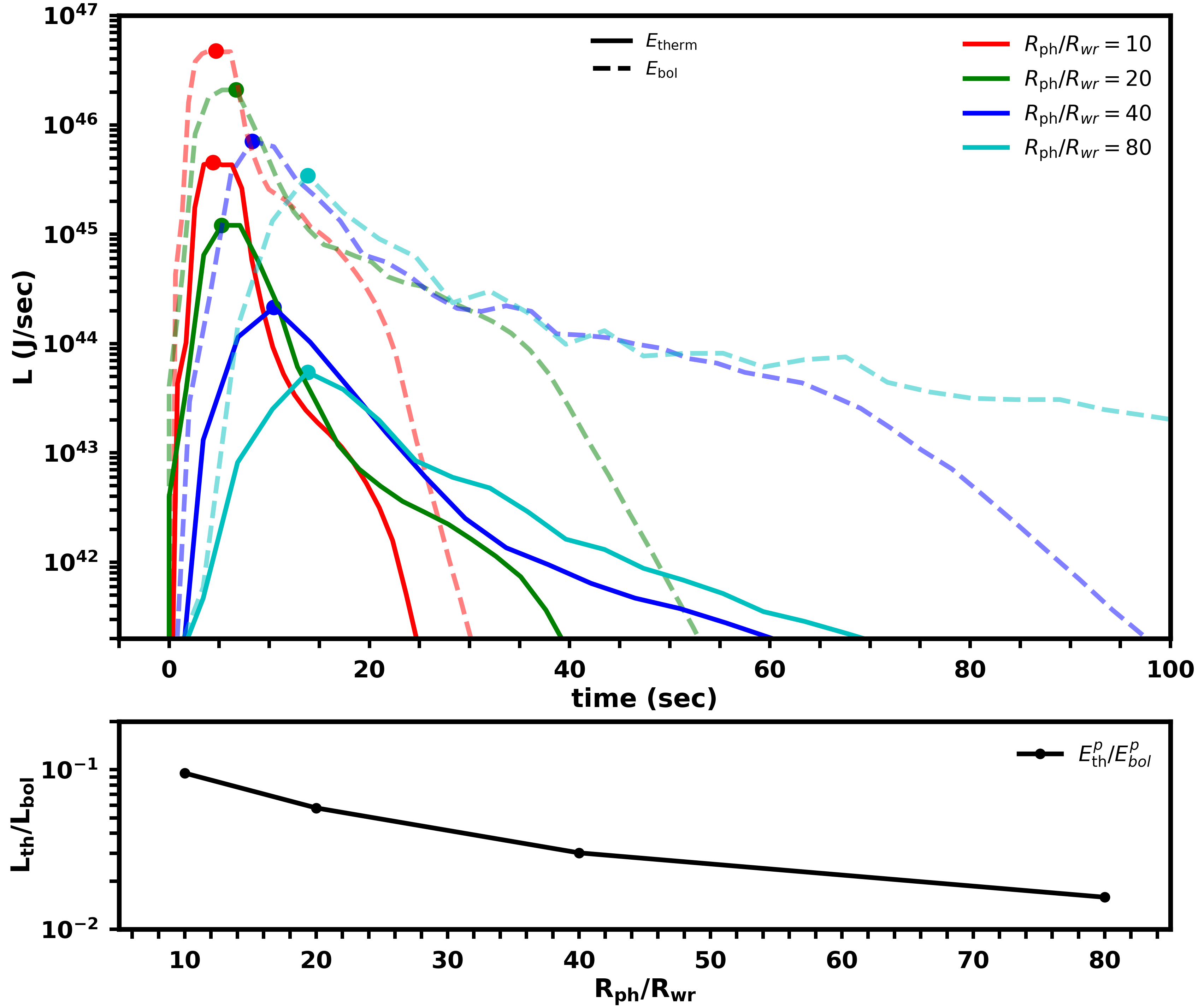}
\caption{\footnotesize{Comparison for thermal (solid lines) and bolometric (dashed lines) light curves for an sudden shut-off jet (Eq. \ref{Eq:sharp_cutoff}) with $t_{\rm{eng}} = 2.8$ sec $(\sim t_{\rm{bo}})$ at different photospheric radii. The bottom plot shows the fraction of thermal to bolometric luminosity as a function of distance from the stellar surface.}}
\label{fig:lc_bol_term}
\end{figure}

While we consider the bolometric luminosity for our light curves, the GRB emission is dominated by the thermal energy within the jet. The thermal energy light curves for a sudden shut-off jet model (Eq. \ref{Eq:sharp_cutoff}), for given engine duration ($t_{\rm{eng}} \sim t_{\rm{bo}}$) at different photospheric radii, are shown in Fig. \ref{fig:lc_bol_term}. While the observed luminosity values decrease the overall behavior of the prompt emission remains the same.

For a successful jet breakout, we also investigate the imprints of jet structure variability on lGRB lightcurves. Fig. \ref{fig:lightcurves_vj} shows lGRB lightcurve for an input jet profile that varies sinusoidally as sin$(2\pi t/10t_{\rm{eng}})$. We find for $t_{\rm{eng}} \approx 1.8$ sec ($< t_{\rm{bo}}$) with a successful breakout, the lGRB shows no such variability in the bolometric lightcurve even at larger photospheric radii. While for longer engine durations, $t_{\rm{eng}} \approx 232$ sec ($\gg t_{\rm{bo}}$), the jet structure variability is reflected in the lGRB lightcurve. These variations follow the input jet variability in temporal scales with diminished luminosity at all photospheric radii.
\section{Discussion} \label{sec:Discussion}

The results presented above can be summarized in three categories based on the $\left<\Gamma\beta\right>_{L(t)}$, which in turn depends on the engine duration. The first regime, $t_{\rm{eng}} \ll t_{\rm{bo}}$: the engine shuts off while the jet cocoon is still well within the star. Therefore, they have $\left<\Gamma\beta\right>_{L(t)} < 1$. These jets hence fail to produce a gamma ray burst. We thus call this \textit{the failed GRB regime}. The red shaded region marks this in Fig. \ref{fig:t90_raw}.

\begin{figure}
%\flushright
\gridline{\fig{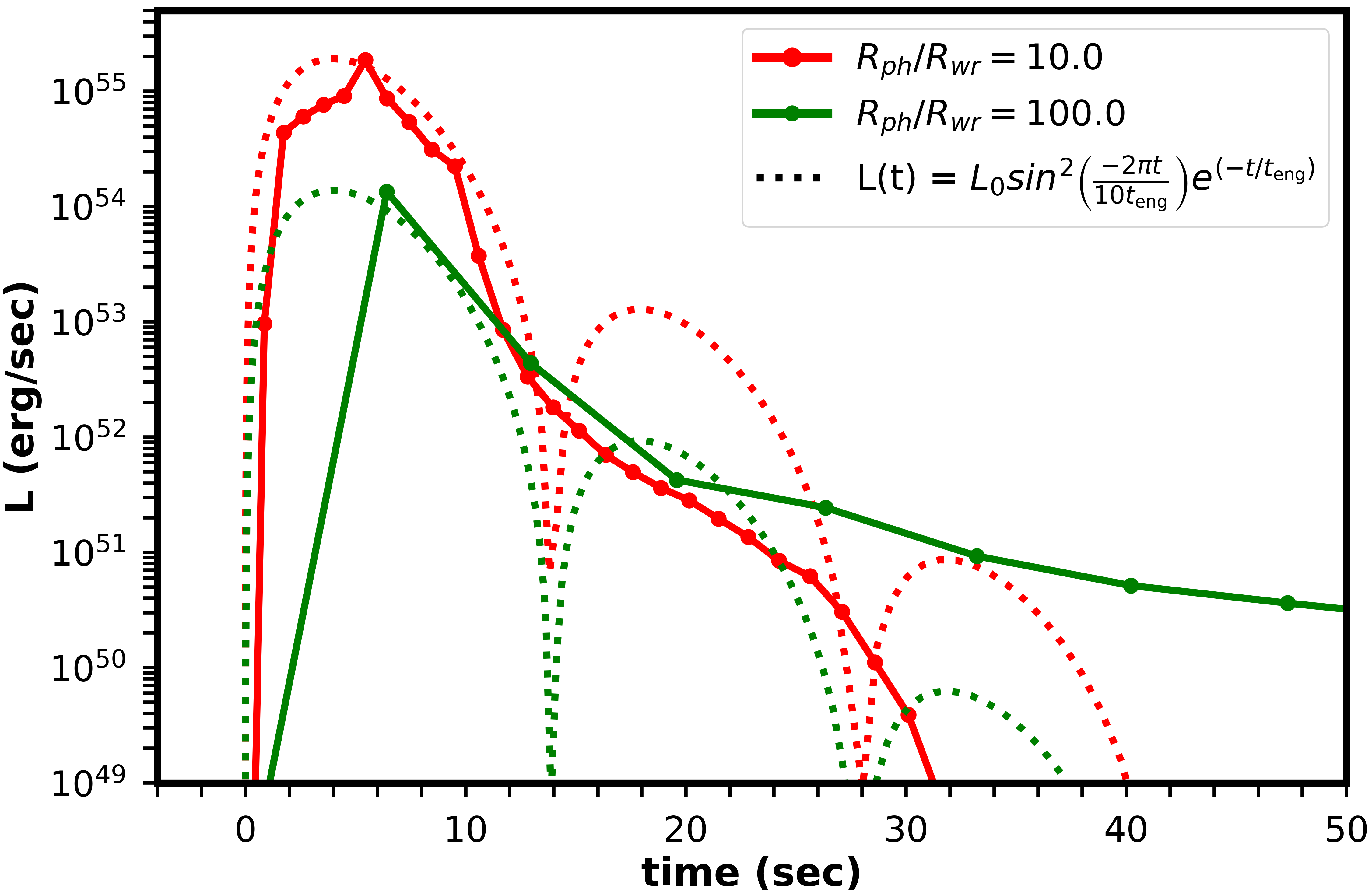}{0.8\linewidth}{\footnotesize{(a) lGRB lightcurve for $t_{\rm{eng}} \approx 2.8$ sec.}}}
    \gridline{\fig{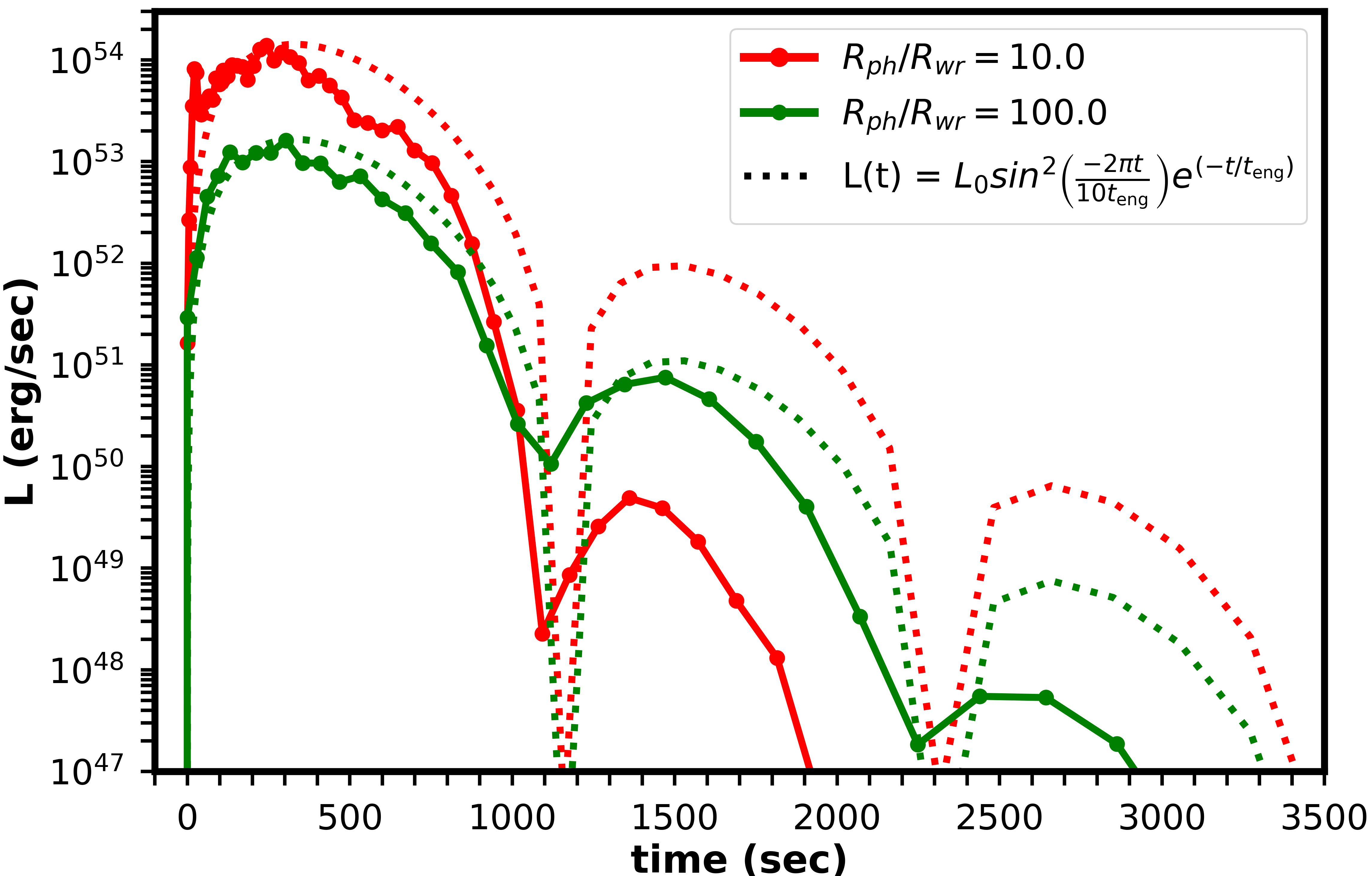}{0.8\linewidth}{\footnotesize{(b) lGRB lightcurve for $t_{\rm{eng}} \approx 232$ sec.}}}
    \caption{\footnotesize{lGRB lightcurve for a sinusoidal jet structure. The dashed lines show the input jet profile, $L_{\rm{eng}}(t) = L_0\times\mbox{sin}^2\left(\frac{-2\pi t}{10t_{\rm{eng}}}\right)\mbox{exp}\left(\frac{-t}{t_{\rm{eng}}}\right)$. The solid lines plot the corresponding lGRB bolometric lightcurves. In (a), since the jet head dominates the emission for a shorter engine duration, the jet variability gets smeared out and we get a smooth profile at all photospheric radii. In (b) for a longer engine duration ($t_{\rm{eng}} > t_{\rm{bo}}$), as the emission is dominated by the unshocked jet, the variabilities within the jet structure are reflected with the same frequency in the lGRB lightcurves. In our model, this is manifested as periodic dips in the lighjtcurve coincident with the input jet profile. All the lightcurves are shifted to begin at $t=0$.}}
    \label{fig:lightcurves_vj}
\end{figure}

The second regime, $t_{\rm{eng}} \sim t_{\rm{bo}}$: the engine is shut off around the time the jet cocoon breaks out of the star. Therefore, given that $\left<\Gamma\beta\right>_{L(t)}$ remains greater than 1, and the lGRB lightcurve reflects the shocked jet head profile. Consequently, $t_{\rm{90}}$ increases as the photosphere shifts further away from the stellar surface. This is a crucial result that shows dependence of $t_{\rm{90}}$ with $R_{\rm{ph}}$. The GRB duration at a given $R_{\rm{ph}}$ also increases in this regime as $t_{\rm{eng}}$ increases. Hence, $t_{\rm{90}}$ also has strong $t_{\rm{eng}}$ dependence. The yellow shaded portion in Fig. \ref{fig:t90_raw} marks this regime. Further, due to the jet head dominating the emission, any variability within the jet structure is smeared out and we get a smooth lGRB bolometric lightcurve from this regime (Fig. \ref{fig:lightcurves_vj}).

\begin{figure*}
\centering
\includegraphics[width=0.8\linewidth]{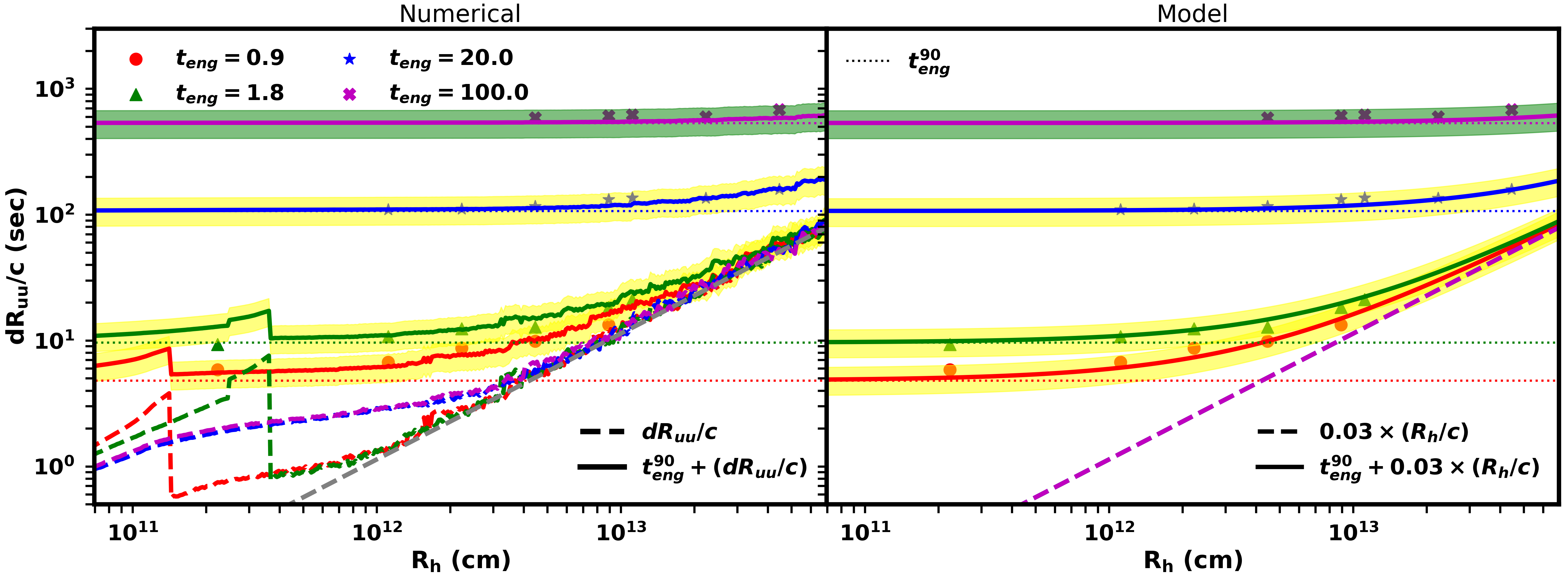}
\caption{\footnotesize{Plot for photospheric dependence of $t_{90}$ for a given $t_{eng}$. The left panel plots our numerical results. The colored dashed line, $dR_{uu}/c$, is the light-crossing time between the shocked jet head and the unshocked jet head, the jet head width for each $t_{eng}$. This is the distance between the white dashed lines for the line plots in Fig. \ref{fig:hydro_evol}. The markers represent the $t_{90}$ (Fig. \ref{fig:t90_raw}) for the given jet. The long-term evolution of $dR_{uu}/c$ can be modeled as $0.03\times(R_{h}/c)$, where $R_{h}$ is the distance of the jet head from the stellar core. The right panel plots the results of our model, $t_{90} = t^{90}_{eng} + 0.03\times(R_{\rm{ph}}/c)$, as a function of jet head distance (effectively $R_{ph}$), keeping $t_{eng}$ fixed. The engine durations are highlighted by the color of the region they belong to in Fig. \ref{fig:t90_raw}. The dotted horizontal lines are the corresponding $t^{90}_{eng}$.}}
\label{fig:t90_rph_model}
\end{figure*}

\begin{figure*}
\centering
\includegraphics[width=0.9\textwidth]{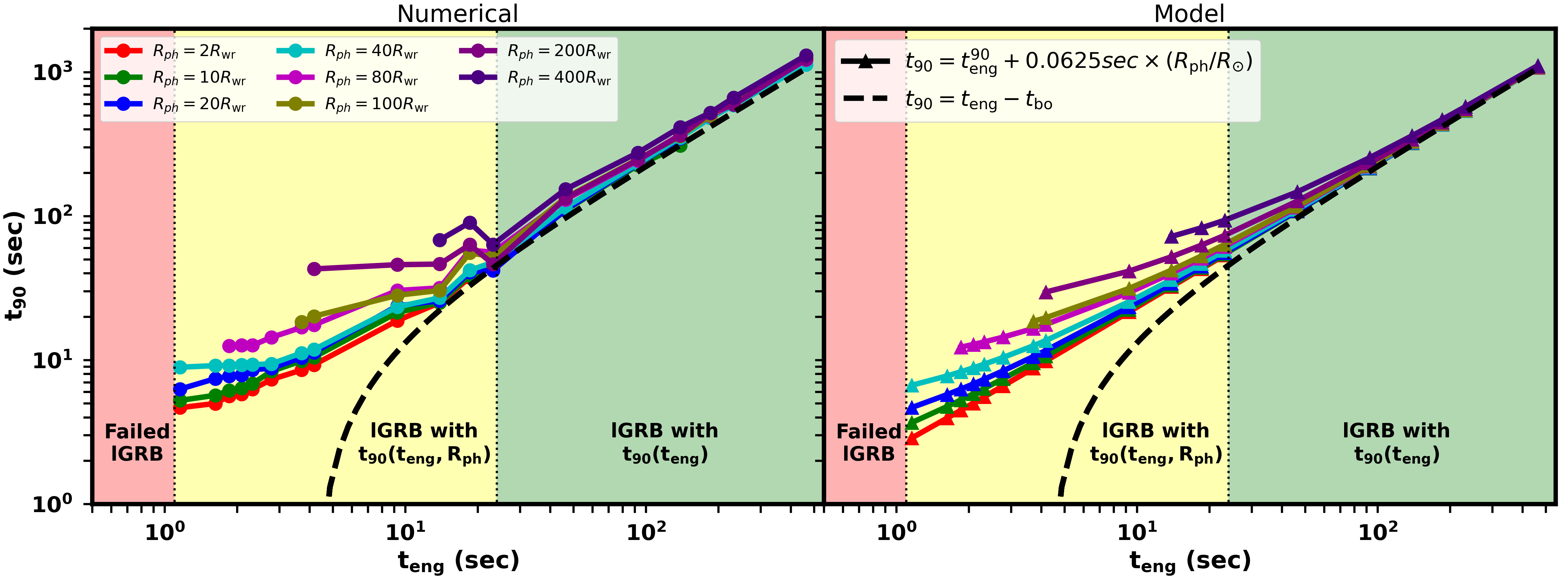}
\caption{\footnotesize{This plot shows the results for our model. The left panel is the same as the lower panel of Fig. \ref{fig:t90_raw}. The right panel plots the results of our model, $t_{90} = t^{90} + 0.07\mbox{ sec }\times(R_{\rm{ph}}/R_{\odot})$, for the corresponding $t_{\rm{eng}}$ and $R_{\rm{ph}}$. The solid black line in the right panel shows the analytical breakout time for the jets (assuming the jet head travels almost at the speed of light). The dashed black line in both panels plots the Bromberg model for comparison.}}
\label{fig:t90_model}
\end{figure*}

The third regime, $t_{\rm{eng}} \gg t_{\rm{bo}}$: the engine shuts off long after the jet has broken out of the star. As a result of this, they break out at ultra-relativistic velocities and continue to remain so for long and $\left<\Gamma\beta\right>_{L(t)}$ saturates to a high value ($\sim 4\times 10^2$). The same is also reflected in the GRB lightcurve which reflects the input profile for the jet (Fig. \ref{fig:t90_raw}). We hence obtain the well known result $t_{\rm{90}} \sim t^{90}_{\rm{eng}}$ in this limit. Where $t^{90}_{\rm{eng}}$ is the duration over which 90\% of the total input energy is deposited in the jet analytically. Finally, this region is marked by the green shaded region in Fig. \ref{fig:t90_raw}. Any variability within the jet structure is also consequently reflected in the lGRB bolometric lightcurve for such jets (Fig. \ref{fig:lightcurves_vj}). The last two regimes, $t_{\rm{eng}} \sim t_{\rm{bo}}$ and $t_{\rm{eng}} \gg t_{\rm{bo}}$ emit a successful prompt lGRB. The $t_{\rm{90}}$ in these regimes depend on the photospheric radius and the engine duration, the effects of which can be modeled together as (Fig. \ref{fig:t90_model}):
\begin{figure}
\centering
\includegraphics[width=0.5\textwidth]{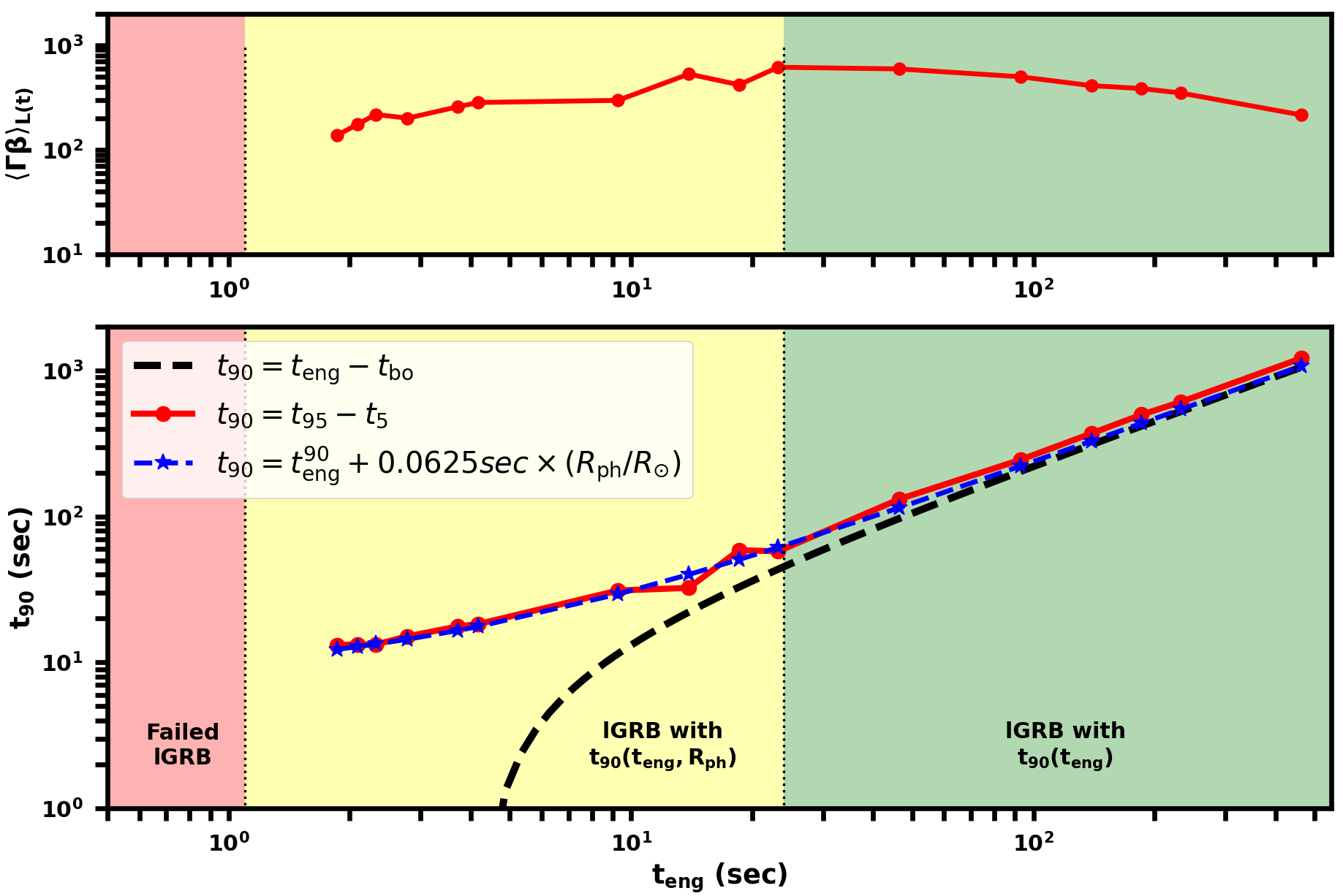}
\caption{\footnotesize{This plot is same as Fig. \ref{fig:t90_raw} at a given photospheric radius $R_{\rm{ph}} = 80 R_{\rm{wr}} = 128 R_{\odot}$. The blue curve in the lower panel plots our $t_{90}$ and $t_{\rm{eng}}$ model over the numerical results. The Bromberg model is also plotted in a black dashed line for comparison. Significant deviation can be seen between the two models for shorter engine durations.}}
\label{fig:t90_gambavg_80}
\end{figure}

\begin{equation}
    t_{\rm{90}} = t^{\rm{90}}_{\rm{eng}} +  0.03\times\left(\frac{R_{\rm{ph}}}{c}\right)
    \label{eq:T90_Model}
\end{equation}

Since the jet head is the dominant contributor to luminosity for a given engine duration, the photospheric dependence of $t_{90}$ can be modeled by tracking the jet head width ($dR_{uu}$) as the jet propagates (Fig. \ref{fig:t90_rph_model}). The width of the jet head can be modeled as $dR_{uu} \approx 0.03(R_h)/c$, where $R_h$ is the distance of the jet head from the collapsed core. Note that the scaling with $R_{\rm{ph}}$ in Eq. \ref{eq:T90_Model} might be modified if the most extremely relativistic material is concentrated within a thinner shell proportional to the inverse square of the Lorentz factor ($1/\Gamma^2$).  However, for such high Lorentz factors, the flow has not yet achieved homology, and therefore we report our results agnostic to what the most extremely relativistic material is doing. The stellar radius dependence can be removed by re-writing the above relation as,

\begin{equation}
    t_{\rm{90}} = t^{\rm{90}}_{\rm{eng}} +  0.07\mbox{ sec }\times\left(\frac{R_{\rm{ph}}}{R_{\odot}}\right)
    \label{eq:T90_Model_2}
\end{equation}

Fig. \ref{fig:t90_gambavg_80} over plots our model (blue dashed line) (Eq. \ref{eq:T90_Model}) with our results (red solid line) at $R_{\rm{ph}} = 100 R_{\rm{wr}}$. The top plot constraints the $t_{\rm{eng}}$ at $R_{\rm{ph}}$ for which a lGRB is emitted, that is $\left<\Gamma\beta\right>_{L(t)} > 1.0$. The bottom plot shows the $t_{\rm{eng}} - t_{\rm{90}}$ correlation. We compare this with the known relation $t_{\rm{90}} = t_{\rm{eng}}-t_{\rm{bo}}$, plotted in gray dashed line in the lower panel. Fig. \ref{fig:t90_model} compares our results and our model at different photospheric radii. This plot shows only the successful lGRB emitted. That is, the emission from $t_{\rm{eng}} \ll t_{\rm{bo}}$ with $\left<\Gamma\beta\right>_{L(t)} < 1.0$ are not included, along with jets which are in the decelerating phase for a given photosphere. In addition, due to numerical limitations in resolving the domain scales, longer jets at smaller photospheric radii had to be manually omitted. The relation Eq. \ref{eq:T90_Model} holds even at larger photospheric radii.

\begin{figure}
\centering
\includegraphics[width=\linewidth]{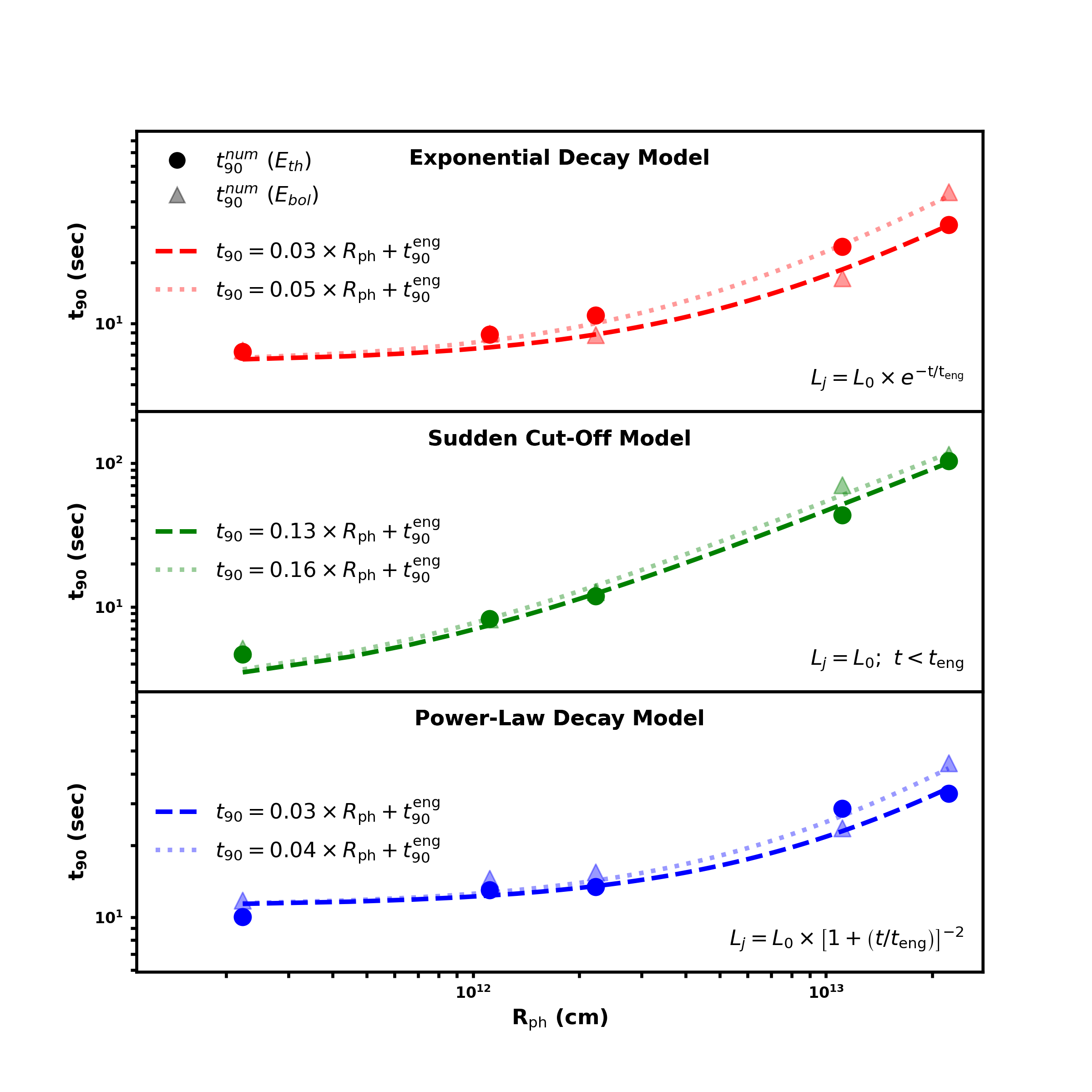}
\caption{\footnotesize{The dependence of $t_{90}$ on the time-dependence of the engine model. The dashed lines correspond to correlations obtained for thermal energy light curves, while the dotted lines correspond to bolometric light curves. The three different models in Eq. \ref{Eq:cutoffs}, are plotted in the three panels (top: Eq. \ref{Eq:exp_cutoff}; middle: Eq. \ref{Eq:sharp_cutoff}; bottom: Eq. \ref{Eq:square_cutoff}). The $t_{90}$ obtained from the lightcurves at $R_{\rm{ph}}/R_{\rm{wr}} = {2, 10, 20, 100, 200}$ for an engine duration of $t_{\rm{eng}} \sim 2.8$ sec ($< t_{\rm{bo}}$) are plotted in solid circles. All of the three jet cut-off models follow the same correlation $t_{90} = \alpha R_{\rm{ph}} + t_{90}^{\rm{eng}}$. Where, the parameter $\alpha$ is model dependent. We consider only the relativistic tail of the power law jet model, which is 80\% of total energy.}}
\label{fig:t90_teng_models}
\end{figure}

Similarly, $t_{90}$ for lightcurves from jets following models Eq. \ref{Eq:sharp_cutoff} and Eq. \ref{Eq:square_cutoff} are tracked as a function of the jet evolution. The results are presented in Fig. \ref{fig:t90_teng_models}. Regardless of the jet model, we find an additional photospheric-dependent component contributing to the $t_{90}$. This extends the lGRB duration beyond the intrinsic $t_{90}^{\rm{eng}}$ expected for that model, especially for shorter engine durations. The model from Eq. \ref{eq:T90_Model} can then be generalized for all the jet models as:

\begin{equation}
    t_{90} = t^{90}_{\rm{eng}} + \alpha\times\left(\frac{R_{\rm{ph}}}{c}\right)
    \label{eq:T90_model_general}
\end{equation}

Here, the parameter $\alpha$ depends on the jet cut-off model and jet parameters

\section{Conclusion}  \label{sec:Conclusion}

In this work, we investigate the correlation of central engine activity duration ($t_{\rm{eng}}$), powering a jet, of a collapsed Wolf Rayet star and the lGRB duration ($t_{\rm{90}}$) (if successful). The jet traverses through the remaining stellar envelope Eq. (\ref{Eq:Density_WR}), and given enough energy breaks out of the star ($t \equiv t_{\rm{bo}}$). Following a successful breakout, we assume a photospheric GRB emission at $R_{\rm{ph}}$. For each jet, we calculate the luminosity weighted average four velocity ($\left<\Gamma\beta\right>_{L(t)}$), given by Eq. \ref{eq:lum_avg_four_velo}, to determine the success of lGRB emission at a given $R_{\rm{ph}}$. Since lGRBs are emitted by the (ultra-)relativistic and luminous material, if $\left<\Gamma\beta\right>_{L(t)} > 1$ we categorize them as lGRB candidates. Otherwise, if $\left<\Gamma\beta\right>_{L(t)} \leqslant 1$ they are categorized as lGRB-less failed jet supernova. For each jet with a given $t_{\rm{eng}}$, we generate a bolometric lightcurve (Fig. \ref{fig:lightcurves}) and estimate the $t_{\rm{90}}$ from it. Irrespective of $t_{\rm{eng}}$, jets that successfully break out of the star do so with $t_{\rm{bo}} \sim 4.3$ seconds for our jet and stellar parameters (Fig. \ref{fig:t90_raw}).

As discussed in Sec. \ref{sec:Results}, our results show three distinctive categories of lGRB based on their $t_{\rm{90}}$ (Fig. \ref{fig:t90_raw}). The first category is that of failed jets (red-shaded region), that is jets that do not emit a GRB. We find the upper limit for central engine duration in this category as, $t_{\rm{eng}} \leqslant 1.2$ seconds ($ \sim 0.3 t_{\rm{bo}}$). These jets have engine duration significantly shorter than the breakout time, that is $t_{\rm{eng}} \ll t_{\rm{bo}}$. The second category of jets (marked by the yellow region in Fig. \ref{fig:t90_raw}), shows significant photospheric and $t_{\rm{eng}}$ dependence on the $t_{\rm{90}}$. The region spans over $t_{\rm{eng}} \in (1.2-24)$ seconds, which corresponds to $\sim 0.3 t_{\rm{bo}} - 5.6 t_{\rm{bo}}$. Hence, this region is marked by $t_{\rm{eng}} \sim t_{\rm{bo}}$. The $t_{\rm{90}}$ is dictated by the light crossing time in the jet head at any given photospheric radii. The lower limit, $t_{90,min} = t_{\rm{eng,min}} + 0.07\mbox{ sec }\times(R_{\rm{ph,min}}/R_{\odot}) \approx 1.4$ seconds, in this region serves as the shortest possible $t_{\rm{90}}$ in our model. Finally, the third category of jets (marked by the green region), have lGRB duration dictated solely by $t_{\rm{eng}}$. Jets with $t_{\rm{eng}} > 24 \mbox{ seconds } (\sim 5.6 t_{\rm{bo}})$ fall under this category. This region is hence marked by $t_{\rm{eng}} \gg t_{\rm{bo}}$. So, the longest possible $t_{\rm{90}}$ is dictated by the longest possible time for which the central engine can power the jet.

We thus propose a new model for lGRB $t_{\rm{90}} = t^{90}_{\rm{eng}} + \alpha\times (R_{\rm{ph}}/c)$ (Eq. \ref{eq:T90_model_general}). For our parameters of a WR star, with stellar radius $R_{\rm{wr}} = 1.6R_{\odot}$ and $M_* = 4 M_{\odot}$. This can equally be scaled in terms of the stellar radius as, $t_{\rm{90}} = t^{90}_{\rm{eng}} + \left(\frac{\alpha R_{\odot}}{c}\right)\mbox{ sec }\times(R_{\rm{ph}}/R_{\odot})$. However, the caveat here is that this coefficient $\alpha$ also depends on the jet properties such as power, opening angle, and the detailed time-dependence of the engine.  We plan to explore the effects of jet properties on this term in follow-up studies. A similar treatment for thermal energy GRB light curves show the same trend. However, due the conversion of thermal to kinetic energy the coefficient of of expansion with the photospheric radius is smaller (Fig. \ref{fig:t90_teng_models}).

Now let's look at some of the direct consequences of this model. Traditionally in the Bromberg model (\citep{Bromberg+2012}), long gamma ray burst jets are categorized based on successful breakout ($t_{\rm{eng}} > t_{\rm{bo}}$) or not ($t_{\rm{eng}} < t_{\rm{bo}}$) (corresponding failed jets). For comparison this is plotted as the black dashed line in Fig. \ref{fig:t90_model}. As expected, the Bromberg model has a sharp cut-off on $t_{\rm{90}} \sim t_{\rm{bo}}$, which we do not find. Thus, in the engine duration regime, $t_{\rm{eng}} > 1.2$ sec to $t_{\rm{eng}} \approx 4.3$ $(\sim t_{\rm{bo}})$ we get successful jet breakout and a GRB consequently. In contrast to the standard GRB duration correlation, which predicts a failed jet breakout in this regime. Further, within the yellow regime, we predict an additional monotonically increasing dependence of $t_{\rm{90}}$ on the photospheric radius. While in the standard picture, the only dependence is on the difference $t_{\rm{eng}} - t_{\rm{bo}}$. For yet longer engine duration, $t_{\rm{eng}} \gg t_{\rm{bo}}$, our model predicts the same $t_{\rm{90}}$ as the Bromberg model, that is $t_{\rm{90}} \sim t_{\rm{eng}}$. This also shows that the shortest duration long GRBs should not be variable in nature. This may differ for short engine duration and very high luminosity jets, where internal shock collisions can occur before the jet breakout. This should be especially relevant for llGRBs, which are typically not variable in nature (\cite{Nakar+2012, Nakar+2015}). In addition, variabilities within the jet structure are reflected in the lightcurve for jets which shuts off after breaking out of the stellar surface.

This regime also predicts a minimum possible value of $t_{\rm{90}}$, which can be readily seen in Equation \ref{eq:T90_Model_2} or Figure \ref{fig:t90_model} to be a few seconds long.  Of course, this minimum value will depend on the precise coefficient $\alpha$ in Equation \ref{eq:T90_model_general}, which may depend for example on jet luminosity, opening angle, or progenitor radius.  Future work should investigate these dependencies further, to determine a true theoretical minimum that can be tested against observations.

\cite{Bromberg+2011} analyzed a set of 5 llGRBs and estimated the breakout times based on luminosity and assumed stellar properties and a jet opening angle. They found three out of the five llGRBs, GRB980425/1998bw, GRB031203/2003lw, and GRB051109B have $t_{\rm{90}}$ greater than the inferred engine duration. In the standard picture, this could be explained by geometric effects (\cite{Paz+2020}). However,  our model provides an interpretation for the duration of such low-luminous lGRBs where the extended $t_{\rm{90}}$ duration arises from larger photospheric radii for a short engine duration. These lGRBs also do not show variability in their lightcurves, as expected for short-engine duration jets in our model.

The correlation of lGRB duration with engine duration also serves as a model to estimate the lGRB population distribution \citep{Bromberg+2012}. In our upcoming works, we plan to use our new model to estimate the population distribution of lGRB, and investigate the low luminous lGRB population density in detail. A re-evaluation of archival data from \textit{Swift} and \textit{Fermi} lGRB observations may further reveal photospheric dependence for existing observations. We further plan to extend a similar analysis for short Gamma Ray Bursts and investigate similar correlations. Upcoming missions with enhanced onboard GRB detection like the \textit{Space Variable Objects Monitor} (SVOM) and \textit{BurstCube} will further enhance the lGRB and sGRB catalog. Many aspects of astronomical Gamma Ray Burts remain debated. The GRB duration remains one of the best parameters to categorize them. Our work provides a new model basis for this categorization in the light of new and exotic GRB detection.

\begin{acknowledgments}

We acknowledge C. M. Irwin, R. Perna, and W. Fong for their insightful suggestions and for bringing to our notice anomalous GRB observations. We also acknowledge D. Giannios and D. Milisavljevic for their expert comments and discussions on the subject. Hydrodynamical calculations were carried out on the Petunia cluster at Purdue University.  This project and the development of the code \textit{Firefly} was supported by NASA under grant No. 80NSSC22K1615.
\end{acknowledgments}

\typeout{}
\bibliography{rana}

\end{document}